\documentclass[12pt]{iopart}
\usepackage{caption}
\usepackage{graphicx}
\usepackage{xcolor}
\usepackage{wasysym}
\begin{document}

\title[Magneto-optical trapping in a near-surface borehole]{Magneto-optical trapping in a near-surface borehole}

\author{Jamie Vovrosh$^{1}$, Katie Wilkinson$^{1}$, Sam Hedges$^{1}$, Kieran McGovern$^{1}$, Farzad Hayati$^{1}$, Christopher Carson$^{2}$, Adam Selyem$^{2}$, Jonathan Winch$^{1}$, Ben Stray$^{1}$, Luuk Earl$^{1}$, Maxwell Hamerow$^{1}$, Georgia Wilson$^{1}$,  Adam Seedat$^{1}$, Sanaz Roshanmanesh$^{1}$, Kai Bongs$^{1}$, Michael Holynski$^{1}$}

%John Tellam$^{3}$,

%Kevin Ridley$^{1}$, Geoffrey de Villiers$^{1}$, Matt Guy$^{3}$

\address{$^{1}$School of Physics and Astronomy, University of Birmingham, Birmingham, B15 2TT, UK}
\address{$^{2}$Fraunhofer Centre for Applied Photonics, Fraunhofer UK Research Ltd., Glasgow, G1 1RD UK}
%\address{$^{3}$School of Geography, Earth and Environmental Sciences, University of Birmingham, Birmingham, B15 2TT, UK}
%\address{$^{3}$Geomatrix Earth Science Ltd, Leighton Buzzard, LU7 4TZ}

\ead{J.A.Vovrosh@bham.ac.uk}
\ead{M.Holynski@bham.ac.uk}
\vspace{10pt}
%\begin{indented}
%\item[]August 2017
%\end{indented}

\begin{abstract}
%The outstanding performance achieved by quantum technology gravity sensors based on atom interferometry has the potential to offer a step-change in our remote sensing capabilities across a range of applications, for example, enabling long-term environmental monitoring through low drift measurements. While ground-based sensors have been demonstrated in real world environments, significant improvements in robustness and reductions to radial size, weight, and power consumption are required for such devices to be deployed in boreholes. To realize the first step towards the deployment of cold atom-based sensors down boreholes, we demonstrate a borehole-deployable magneto-optical trap, the core package of many cold atom-based systems. This demonstrator was used to generate atom clouds at 1 m intervals in a 50 m deep borehole, in an approach similar to how in-borehole gravity surveys are performed. During the trial the system generated on average clouds of (7.3 $\pm 0.1) \times 10^{6}$ $^{87}$Rb atoms with the standard deviation in atom number across the survey observed to be as low as $6.5 \times 10^{5}$. 

Borehole gravity sensing can be used in a number of applications to measure features around a well including rock-type change mapping and determination of reservoir porosity. Quantum technology gravity sensors based on atom interferometry have the ability to offer increased survey speeds and reduced need for calibration. While surface sensors have been demonstrated in real world environments, significant improvements in robustness and reductions to radial size, weight, and power consumption are required for such devices to be deployed in boreholes. To realise the first step towards the deployment of cold atom-based sensors down boreholes, we demonstrate a borehole-deployable magneto-optical trap, the core package of many cold atom-based systems. The enclosure containing the magneto-optical trap itself had an outer radius of ($60\pm0.1$) mm at its widest point and a length of ($890\pm5$) mm. This system was used to generate atom clouds at 1 m intervals in a 14 cm wide, 50 m deep borehole, to simulate an in-borehole gravity surveys are performed. During the survey the system generated on average clouds of (3.0 $\pm 0.1) \times 10^{5}$ $^{87}$Rb atoms with the standard deviation in atom number across the survey observed to be as low as $9 \times 10^{4}$. 

\end{abstract}

%\maketitle

\section{Introduction}

Gravity surveys are used to detect features through their density contrasts, and have an advantage over alternative techniques in that gravity is not attenuated by the intervening medium, such as borehole casing. In comparison with techniques such as ground penetrating radar or nuclear logging, this can allow for the detection of features at greater distances \cite{BODDICE2017149}. Additionally gravity sensing avoids the use of radioactive isotopes \cite{Logging} mitigating security and health concerns. These benefits have led to gravity surveys being used in a number of environments to detect features of interest \cite{doi:10.1190/1.1445845}, including below the surface in boreholes. Borehole gravity sensing is used for a number of applications, including: remote sensing of gas and oil zones behind casing \cite{CHO2020107054}; vertical density profiling for gravity map interpretation and for seismic modelling and analysis \cite{LINES1991183}; detection of geologic structures \cite{osti_6707051,doi:10.1190/1.1442646}; determination of reservoir porosity for reserve estimates \cite{beyer1978density}; monitoring of reservoir fluid conditions for production evaluation, and rock-type change mapping \cite{10.1306/03B5ACE6-16D1-11D7-8645000102C1865D,lafehr1983rock} for groundwater and engineering studies; Carbon Capture and Storage (CCS) monitoring \cite{APPRIOU2020102956,DODDS2013421}; as well as tests of fundamental physics \cite{PhysRevLett.65.1173,PhysRevLett.62.985}.

Despite numerous applications for borehole gravity sensing, a major drawback is measurement time when compared to other remote sensing techniques \cite{doi:10.1098/rsta.2016.0238}. This is primarily due to the need to average out micro-seismic vibrations and calibrate for the inherent drift in existing sensors, via repeat measurements of the same point between sets of measurements. This limits its widespread use as a measurement technique. Quantum technology gravity and gravity gradient sensors based on atom interferometry \cite{PhysRevLett.67.181,Peters1999} have the potential to overcome these issues \cite{Bongs2019}. Moreover, gravity gradient sensors have the additional benefit of negating issues arising from micro-seismic noise \cite{Stray2021}. 

Quantum technology based on atom interferometry has proven to be a powerful method for precision gravity sensing \cite{Peters1999,Bongs2019,PhysRevLett.111.083001,Peters_2001}, with sensitivities of 4.2 ng/$\sqrt{\textnormal{Hz}}$ having been achieved in existing sensors \cite{PhysRevA.88.043610}. Within laboratories, atom interferometry has enabled precise measurements of the equivalence principle \cite{PhysRevLett.125.191101}, the fine-structure constant \cite{morel:hal-03107990,PMID:29650669}, and the gravitational constant \cite{Rosi2014quantum}. The demonstrated performance of laboratory systems and their inherent low drift has resulted in the development of a number of transportable atom interferometry systems targeting multiple  applications and operational conditions \cite{Bongs2019, MarineGravimetry, space, doi:10.1038/s41598-018-30608-1,Wueaax0800, AirborneGravimetry, Guo2021, Stray2021, atoms10010032,s22166172,doi:10.1126/sciadv.add3854}. 

Before atom-interferometry based-gravity sensors can be deployed down boreholes, a number of challenges need to be overcome. Borehole sensors have strict requirements on size, weight, and power consumption (SWaP), particularly for the radial size of the sensor. Existing borehole sensors typically have a radial diameter of between 10 mm – 200 mm, usually set by the depth of operation. In addition to meeting the required SWaP, sensors need to be robust against environmental and operating conditions. Typically, the environmental conditions and operating requirements become more demanding the deeper the borehole. For example, in the deepest boreholes of up to 12 km \cite{kozlovsky1987superdeep}, sensors can be required to operate at pressures of up to 1.4 kbar, temperatures of up to 200 $^{\circ}$C and at angles from the vertical of up to 90$^{\circ}$. However for the majority of boreholes environmental conditions are much more hospitable, particularly in near surface boreholes.

In this article, for the first time, we demonstrate a compact magneto-optical trap (MOT). A MOT uses light that is detuned slightly below an atomic resonance in conjunction with a quadrupolar magnetic field, to cool and trap clouds of atoms \cite{Steane:92}. The compact MOT system has the required SWaP and environmental robustness to be deployed in a borehole. This demonstrator was used to perform a simulated survey down a 14 cm diameter, 50 m deep, borehole where the water level  was to within a few meters of the ground surface, to assess its robustness under trial conditions. During the deployment the system was able to, on average, produce clouds of ($3.0 \pm 0.1) \times 10^{5}$ $^{87}$Rb atoms. MOTs form a core part of a number of cold atom-based devices including atomic clocks \cite{diddams2001optical}, accelerometers \cite{doi:10.1080/23746149.2021.1946426} and magnetometers \cite{doi:10.1063/1.4803684}. As such, achieving a MOT meeting the form factor and robustness for in-borehole operation is the first step towards achieving a borehole-deployable cold atom-based sensor.

\section{Borehole gravity surveys}

\begin{figure}
\centering
\includegraphics[scale=1]{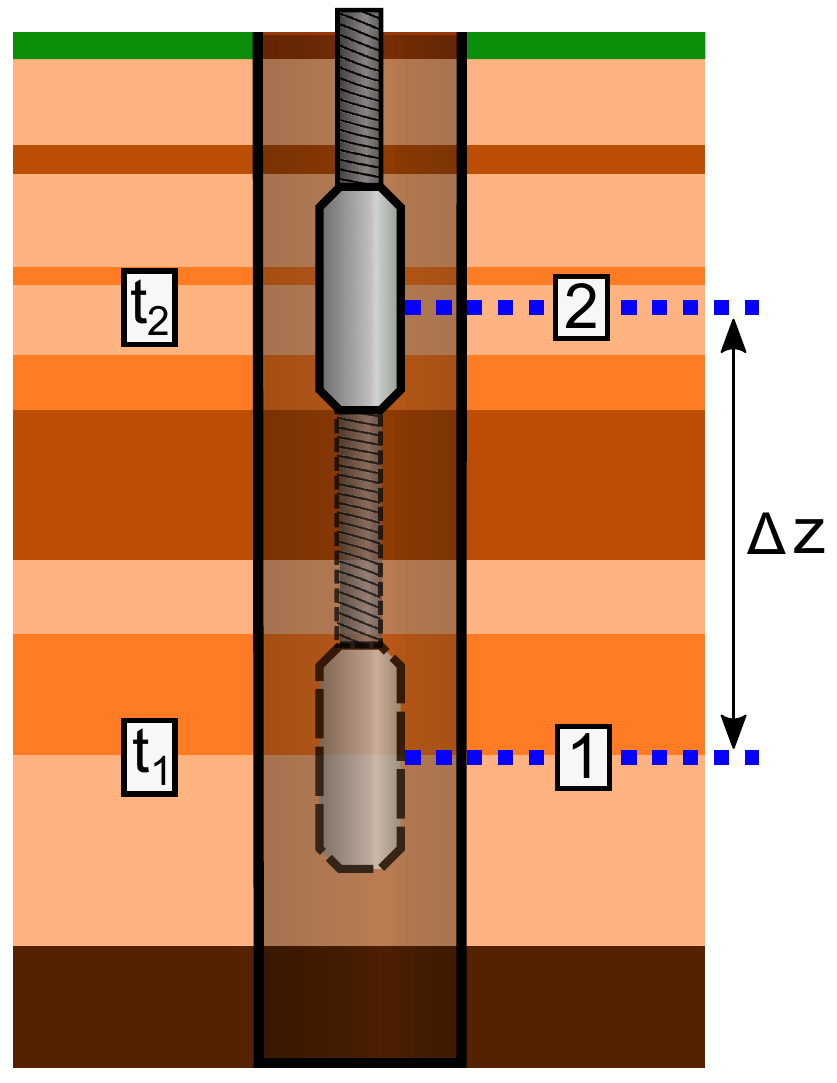}
\caption{In typical borehole gravity sensing, the sensor is lowered or raised to measurements points (denoted as 1 and 2) at fixed distances apart and measurements are performed when the sensor is stationary, with the data collected sequentially in time (t$_{i}$, where $i=1,2$). The subsurface layers are typically assumed to be infinitely extended horizontal slabs of different densities.}
\label{fig:figure 0}
\end{figure}

Typically, borehole gravity measurements are collected at discrete intervals by stopping a deployed gravity sensor at predetermined observation depths and performing a stationary measurement \cite{GRAVILOG}, before moving to the next observation depth, as illustrated in figure  \ref{fig:figure 0}. From these measurements, the vertical gradient of gravity, $\Delta g/\Delta z$, is determined for the interval of interest by measuring the gravity difference, $\Delta g$, and the vertical distance between two consecutive stations, $\Delta z$. 

Existing borehole gravity sensors used to perform such measurements typically rely on spring-based technology similar to that used in surface-based gravimeters \cite{GRAVILOG}. To allow for operation down boreholes the technology used in surface-based sensors has been miniaturized, equipped with self-leveling capabilities and packaged to fit in narrow diameter borehole tools. For example, the sensor discussed in reference \cite{GRAVILOG} has a radial diameter of 57 mm, can be used to depths of 2,500 m and its self leveling capabilities allow it to be deployed in boreholes inclined up to $30^{\circ}$ from vertical. It is typically deployed by conducting wireline cable \cite{GRAVILOG}. Atom interferometry based sensors will need to reach a similar or greater level of deployability to be a competitive tool across all of the applications in which gravity sensing is currently used.

To date, portable atom interferometry based gravity and gravity gradient sensors target surface-based applications and are consequently unsuitable for borehole deployment partly due to radial size and form factor. Several quantum technology-based gravity sensors comprise of a separate control system and sensor head \cite{Stray2021,doi:10.1038/s41598-018-30608-1}. The control system typically contains the majority of electronics required to run the system, while the sensor head is the part of the system where the measurements take place. These two components are connected via an umbilical, carrying optical and electronic signals. An early implementation could require that in-borehole atom interferometers have a control system on the surface. While this approach will work for shallow boreholes, it is expected that at certain length of the umbilical, systematics associated with having long optical fibres become a limitation (i.e optical power attenuation). As such, a control system capable of being deployed down a borehole will likely be required for operation at much greater depths. Initial implementations of quantum sensors are likely to operate in a similar manner to the existing borehole gravity sensors, stopping at intervals to perform measurements, with the long-term goal of performing measurements while moving \cite{9430461}, such that it can be used as a continuous logging tool. It is also expected that future quantum gravity sensors in borehole would be able to determine the gradient directly \cite{PhysRevLett.81.971,PhysRevA.65.033608}, reducing errors associated with relative positioning between measurements \cite{GRAVILOG}.

%Performing measurements like this means any error in the recorded depth between two measurement positions will lead to errors in the density measured. For example, an error of 2 cm between measurements using a single sensor at two different positions will result in an error in density of about 73 kg/m$^3$, due to inaccuracy in the free air correction alone\cite{osti_1773345}. Quantum sensors have the potential to address this error due to their ability to generate two interferometers using a single laser beam. For example in reference \cite{Stray2021} a relative positioning error between the two sensors was measured to be less than 75 $\mu$m for a separation of 1 m, which would correspond to an error of 0.3 kg/m$^3$. 

As such, the demonstrator constructed here was designed to have the  package deployed downhole while the control system remains on the surface. The science package was designed to be deployable via a winch, which was stopped at predefined intervals along the depth of the borehole to perform measurements. This allowed for an assessment of the variability and robustness of the system performance, while performing a simulated but representative survey.

\begin{figure}[h]
\centering
\includegraphics[scale=1]{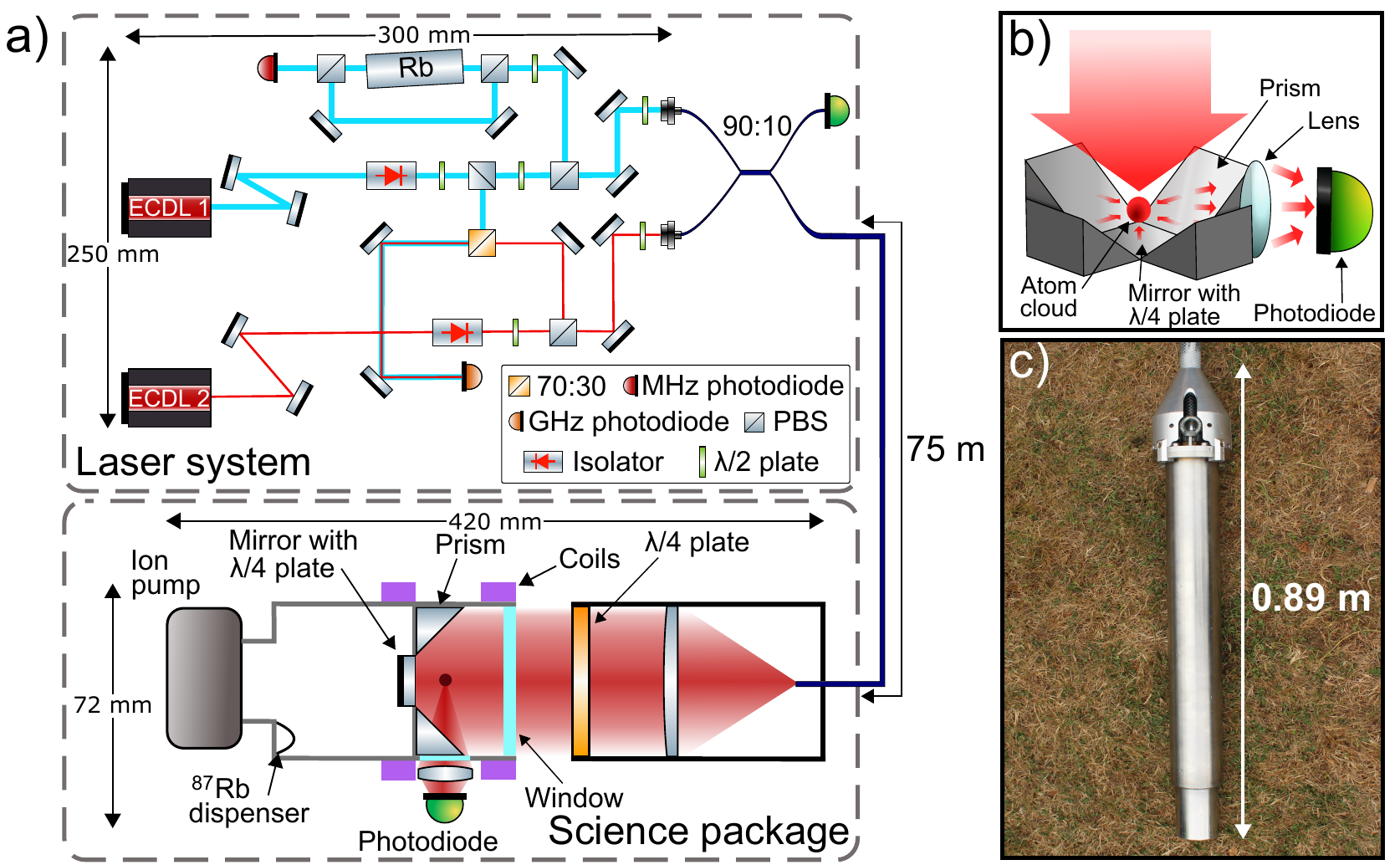}
\caption{ a) Schematic diagram of the laser system and the science package (not to scale). b) Arrangement of the optics used to generate and measure the atom cloud (vacuum walls not shown). c) Photo of the science package. The vacuum system, optics and measurement photodiode are contained within the watertight stainless steel casing shown. The top of the science package is connected to the umbilical, which delivers the light and electrical signals required to operate the science package.}
\label{fig:figure 1}
\end{figure}

\section{Experimental system}

The demonstrator consists of two sub-units (see figure \ref{fig:figure 1} a), a control system and a science package, with optical and electrical signals transferred through a 75 m umbilical. 

The control system primarily consists of a laser system comprised of two Alter UK REMOTE external-cavity diode lasers (ECDLs) \cite{2019SPIE10899E..08D}, each locked to different atomic transitions. The ECDL used to produce cooling light is offset locked \cite{doi:10.1063/1.1149573} to the ECDL used to produce repump light, which is itself stabilised to a rubidium reference via saturated absorption spectroscopy \cite{Lee:96,doi:10.1119/1.16955}. The repump is locked to the $^{87}$Rb $|F=1\rangle$ $\rightarrow$ $|F’=2\rangle$ transition and the cooling laser is offset locked 6.4 GHz from the repump, approximately 10 MHz detuned below the $|F=2\rangle$ $\rightarrow$ $|F’=3\rangle$ transition.  

Optical isolators are placed directly after the output of both ECDLs, providing 35 dB isolation to mitigate unwanted back reflections impacting the frequency performance of the ECDLs. The laser system has a pre-aligned saturated absorption spectroscopy setup where all the optics are bonded in place to provide a robust locking signal to stabilise the repump laser frequency. Separately, a small amount of light from both the cooling and repump beams are split off and overlapped on a fast photodiode to provide a beatnote for the offset lock. Various optics distribute laser light to two fibre couplers where the cooling and repump are combined using a 2x2, 90:10 fibre splitter. The laser system can produce a maximum power of 35 mW of cooling and 5 mW of repump light, greater than the 12.8 mW cooling and 1.9 mW repump light input into the umbilical. The laser system is 250 mm x 300 mm x 80 mm in size, weighs 5 kg and uses robust mechanical mounting techniques to increase portability. 

The light from this laser is delivered to the science package using a 75 m long polarisation-maintaining optical fibre. This fibre delivers the light to a telescope, which produces a circularly polarised beam with 16.6 mm 1/e$^{2}$ diameter and an output power of approximately 8 mW, due to attenuation in the umbilical. This beam is collimated and shone directly into the vacuum chamber in which the four 5 mm $\times$5 mm prisms, a quarter wave plate, and a mirror are used to create the six counter-propagating beams required to cool the atoms in all three degrees of freedom (see figure \ref{fig:figure 1} b) \cite{doi:10.1063/5.0030041}. Coils are used to create a quadrupole magnetic field with a linear gradient and null field at the centre of the trapping region. The atoms are loaded into the trap from background atomic vapour, produced under vacuum with a rubidium dispenser. The scattered light from the atom cloud is measured through fluorescence detection. Vacuum pressure is maintained by a 0.4 ls$^{-1}$ ion pump. The vacuum system, telescope and photo-diode have a form factor of ($420\pm1$) mm $\times$ (72$\pm$1) mm $\diameter$.

The science package is (420$\pm5$) mm long and is contained in a (890$\pm5$) mm long stainless steel waterproof enclosure (see figure \ref{fig:figure 1} c), with an outer radius of (60$\pm0.1$) mm at its widest point and internal radius of (37$\pm0.1$) mm. The end of this enclosure consists of a solid stainless steel block to reduce buoyancy. This casing allows it to be deployed in cased or uncased boreholes and under waterlogged, muddy or dry borehole conditions. The science package is capable of surviving at temperatures of 60 $^{\circ}$C, limited by the glass transition temperature of a few 3D printed polylactic acid parts. Replacing these with aluminium would enable operation at higher temperatures. The science package is expected to be waterproof to pressures of $\approx 100$ bar. The science package does not require centralisation.

%\begin{figure}[h]
%\centering
%\includegraphics[scale=1]{figure 3 - Surface Loading Curve.pdf}
%\caption{Example MOT loading curve taken on the surface. The %loading curve has been fitted with equation 2 from reference %\cite{doi:10.1063/1.4928154}.  The inset shows a photo of an atom %cloud generated in the system, after background subtraction.}
%\label{fig:figure 2}
%\end{figure}

\section{Deployment campaign}

\begin{figure}[h]
\centering
\includegraphics[scale=1]{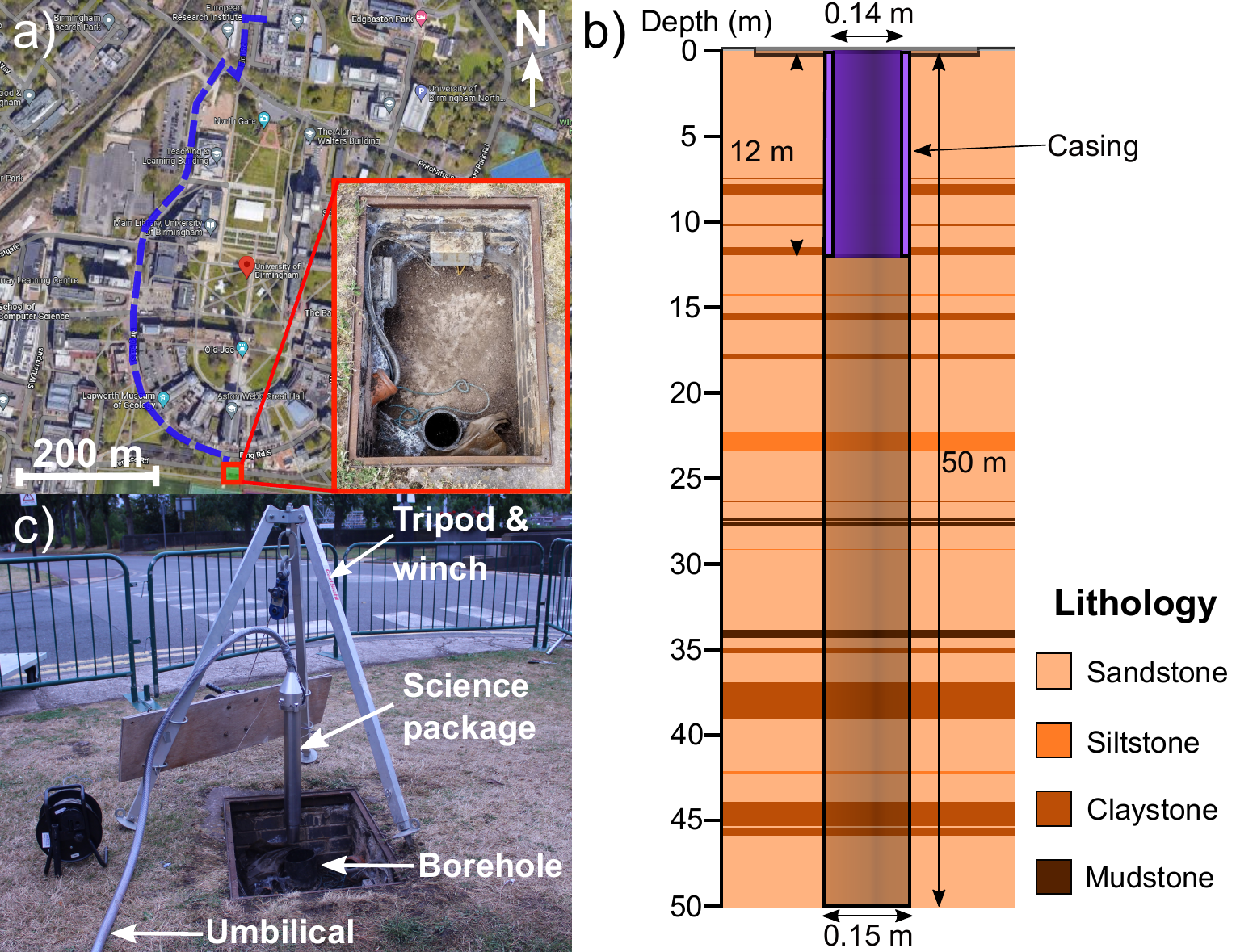}
\caption{a) The location of the test site and route driven to the test site (shown by the blue dashed line). (Map Data: Google, {\small\copyright} 2022, Imagery: Bluesky, Getmapping plc, Infoterra Ltd \& Bluesky, Maxar Technologies, The GeoInformation Group.) The surface of the borehole used for the deployment campaign, is shown in the inset photo (red border). b) Diagram showing the dimensions and lithology of the borehole used in the deployment campaign. The lithological data was obtained from geophysical logging (natural gamma and electrical resistivity), combined with core examination. c) The winch and tripod system used to lower the science package into the borehole. A wire with markings at 1 m intervals was used to record the depth of the device.}
\label{fig:figure 3}
\end{figure}

The system was transported in a medium-sized van, a short distance from the lab to the University of Birmingham campus borehole test site \cite{nerc19549,https://doi.org/10.1029/2010WR009838} for deployment down a borehole as shown in Figure \ref{fig:figure 3} a. During transport the system was unpowered. The borehole itself is 50 m deep and cased to a depth of 12 m with an internal diameter of 14 cm in the cased section (see figure \ref{fig:figure 3} b). The groundwater level was ($3.59\pm0.02$) m below the ground surface level at the time of the survey, determined via the use of a water level meter.

Once at the site, the science package was attached to a cable via the eye holes on the top of the package and lowered into the borehole using a winch and tripod (see figure \ref{fig:figure 3} c). At intervals of (1$\pm0.02$) m below the surface the system was stopped and 5 MOT loading curves were taken. On average about 2 minutes were spent at each depth to record data, limited by the data capture of the oscilloscope used. Moving the system between depths took approximately 4 minutes, limited by the speed of the winch used. 

%Example MOT loading curves taken on the surface and at depths of (25.34$\pm0.02$) m and (49.34$\pm0.02$) m are shown in figure \ref{fig:figure 3.5} a. 

\begin{figure}[h]
\centering
\includegraphics[scale=1]{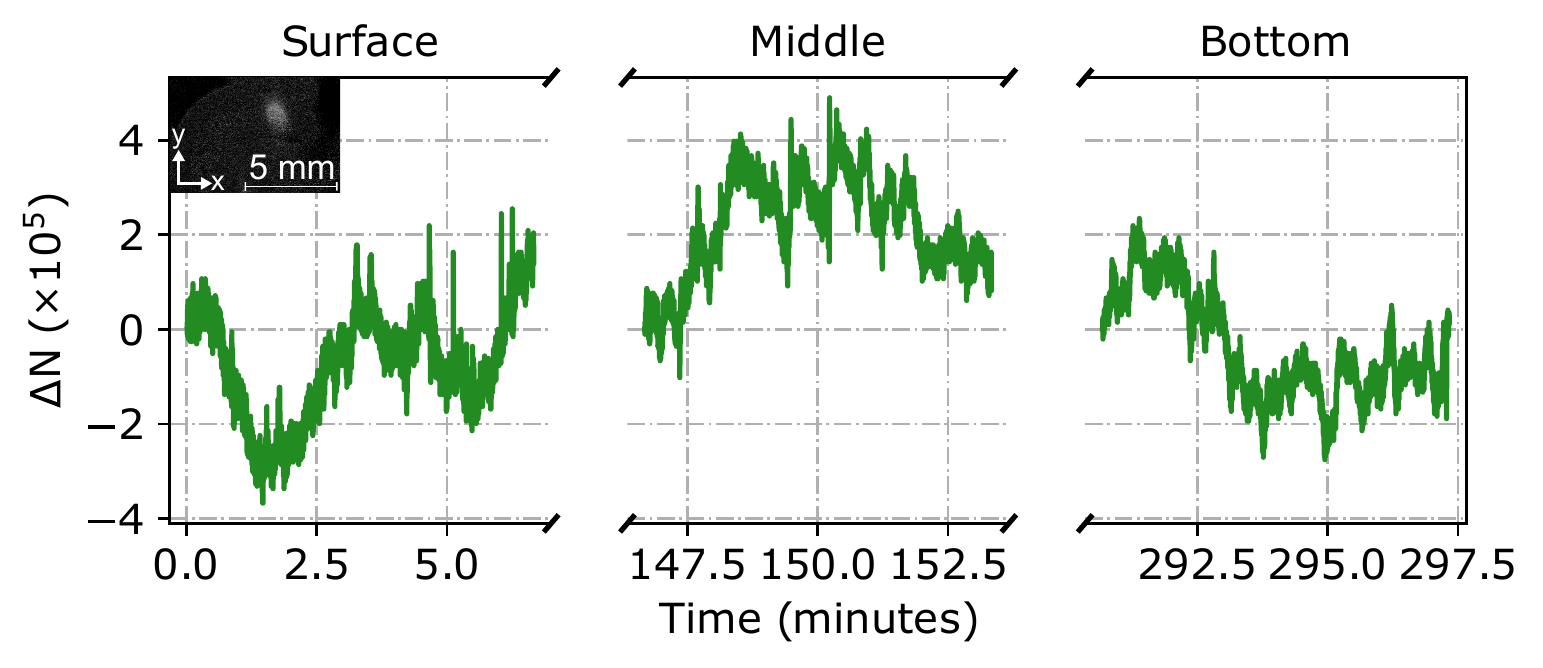}
\caption{The atom number variation with time, relative to the first measurement in each time period, on the surface and at depths of (25.34$\pm0.02$) m and (49.34$\pm0.02$) m in the borehole. The inset photo shows a photo of the cloud on the surface with a radius (1/e$^{2}$) of ($0.81 \pm 0.08$) mm in the x-direction and ($1.1 \pm 0.1$) mm in the y-direction.}
\label{fig:figure 3.5}
\end{figure}

During the survey the stability of the number of atoms in the MOT was measured for 400 s on the surface as well as at the middle, (25.34$\pm0.02$) m and bottom, (49.34 $\pm$ 0.02) m of the borehole. The variation in atom number ($\Delta N$) relative to the start each of time period is shown in figure \ref{fig:figure 3.5}. The standard deviations at the surface, middle and bottom positions were $1.1 \times 10^{5}$, $1.1 \times 10^{5}$ and $1.1 \times 10^{5}$ respectively, suggesting that the MOT stability on short time scales in the borehole and on the surface were comparable. The atom number over the whole deployment had a standard deviation of $8.9 \times 10^{4}$, suggesting that the variation in atom number on short and long time scales was similar.

\begin{figure}[h]
\centering
\includegraphics[scale=1]{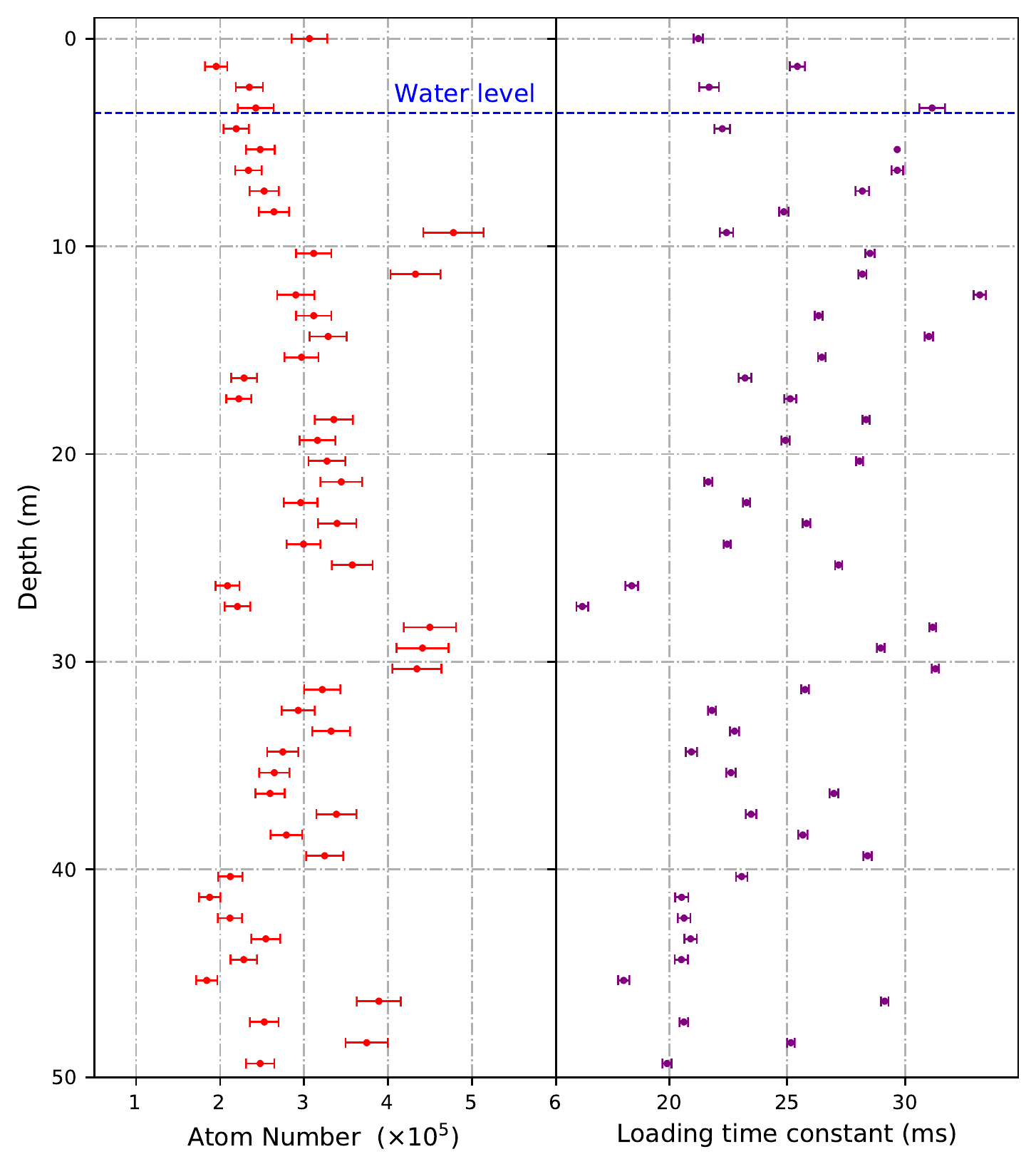}
\caption{MOT parameters as a function of depth in the borehole. The depths were measured with an error of $\pm 0.02$ m.}
\label{fig:figure 4}
\end{figure}

The atom number and loading time constant averaged over 5 curves at each depth can be seen in figure \ref{fig:figure 4}. The loading time constant over the course of the deployment had a standard deviation of 5.00 ms, demonstrating a good level of repeatability. The average atom number achieved in the system over the course of the deployment was ($3.0\pm0.1) \times 10^{5}$, with an average loading time constant of ($24.8\pm0.2$) ms. After reaching a depth of (49.34$\pm0.02$) m and completing all data acquisition, the system was brought back to the surface. The MOT was maintained throughout the trial and was robust to all of the moving and stopping, with no statistically significant effect on the atom cloud observed when becoming submerged under water or exiting the cased part of the borehole. During the trial the pressure in the vacuum chamber varied between $6.5\times10^{-7}$ mbar and $7.7\times10^{-7}$ mbar estimated from the ion pump. This level of variation is typical of performance seen during operation in the lab.

%\begin{figure}[h]
%\centering
%\includegraphics[scale=1]{figure 7 - Stability Data.pdf}
%\caption{Atom number measured over time on the surface and at %depths of (25.34$\pm0.02$) m and (49.34$\pm0.02$) m in the %borehole.}
%\label{fig:figure 4.5}
%\end{figure}

\section{Conclusions and outlook}

The high-precision and low-drift measurements offered by atom interferometry-based gravity sensors, once realised in borehole-deployable packages, have the potential to open up new sensing capabilities, particularly for long-term monitoring applications such as CCS and hydrological monitoring. To push cold atom-based sensors towards the SWaP profile and robustness that is required for deployment and operation in boreholes, the first borehole-deployable cold atom system has been developed and trialed.

The system  was demonstrated to be capable of generating clouds of laser-cooled $^{87}$Rb atoms in a system package of (890$\pm 5$) mm $\times$ (120$\pm1$) mm diameter. The system successfully operated down a 50 m deep borehole, generating atom clouds with an average of (3.0 $\pm 0.1) \times 10^{5}$ atoms. The standard deviation in atom number performance across the survey was as low as $9 \times 10^{4}$. The performance in terms of number of atoms in the trap and atom number stability were comparable to the device performance on the surface and comparable to MOTs used in some existing cold atom sensors \cite{AirborneGravimetry,MarineGravimetry,doi:10.1063/1.4803684}.

To upgrade the demonstrator shown here to a full sensor capable of gravity measurements, several extensions to the system (i.e additional functionality in the laser, magnetic shielding) that meet the requirements for operation down a borehole are necessary. These requirements include meeting sufficiently small SWaP and robustness for deployment (e.g. operation while tilted from the vertical), both of which are active research areas. Examples of innovations which could be utilised to produce SWaP optimised sensors for borehole deployment include SWaP optimised 3D printed components \cite{3DPrint,COOPER2021101898,PRXQuantum.2.030326} and passively pumped vacuum cells \cite{doi:10.1116/5.0053885,rushton2014contributed,burrow2021stand} to allow for operation in shallow boreholes. Further miniaturisation is required to remove the need for a surface based control system  and allow for deployment in deeper boreholes, i.e. compact laser systems \cite{theron2017frequency,luo2019compact,wu2017multiaxis}.

It is also possible that significant improvements in performance over current state of the art laboratory and portable systems may be possible. Trough the incorporation of the latest and future techniques \cite{kovachy2015quantum,PhysRevLett.120.033601,PhysRevLett.111.113002,doi:10.1126/sciadv.abd0650,PhysRevLett.114.063002} and  technological developments \cite{doi:10.1063/5.0030041,doi:10.1063/1.5051663,Hobson2022,PhysRevA.105.L051101}.

\section*{Acknowledgments}

We acknowledge support from EPSRC through grants EP/M013294/1 and EP/T001046/1 and innovate UK through 133991 and 133989, as part of the UK National Quantum Technologies Programme. We would like to acknowledge significant support from John Tellam in regards to planning and carrying out the trial. We would like to acknowledge Alter UK for providing the packaged ECDLs, as well as technical support from the University of Birmingham Engineering and Physical Sciences workshop.

\section*{Data availability statement}
The data that support the findings of this study are available upon reasonable request from the authors.

\section*{Statement of contribution}
The design and development of the demonstrator was performed by J.V, S.H, K.M, K.W, C.C, A.S, F.H, J.W, B.S, L.E, Ma.H, G.W and Mi.H. The survey design and measurements were contributed by J.V, K.W, F.H, J.W, C.C, A.S and S.R. Data processing was carried out by K.W and J.V. J.V, Mi.H, and K.B. conceived and coordinated the experiment. J.V. and K.W. wrote the manuscript. All authors contributed to the review and improvement of the manuscript.

\section*{References}
\bibliography{Ref.bib}

\end{document}